\documentclass[aps,prb,onecolumn,superscriptaddress]{revtex4-1}
\usepackage{graphicx}
\usepackage{bm}
\usepackage{subfigure}
\usepackage{natbib}
\usepackage{lineno}
\usepackage{hyperref}
\usepackage{amsfonts}
\usepackage{amssymb}
\usepackage{courier}

\begin{document}


\title{{\sc Horace}: software for the analysis of data from single crystal spectroscopy experiments at time-of-flight neutron instruments.}

\author{R. A. Ewings}
\affiliation{ISIS Facility,
        STFC Rutherford Appleton Laboratory,
        Harwell Campus, Didcot OX11 0QX, United Kingdom}

\author{A. Buts}
\affiliation{ISIS Facility,
        STFC Rutherford Appleton Laboratory,
        Harwell Campus, Didcot OX11 0QX, United Kingdom}

\author{M. D. Le}
\affiliation{ISIS Facility,
        STFC Rutherford Appleton Laboratory,
        Harwell Campus, Didcot OX11 0QX, United Kingdom}

\author{J. van Duijn}
\affiliation{Departamento de Mec\'{a}nica, Universidad de C\'{o}rdoba, C\'{o}rdoba, 14071, Spain}

\author{I. Bustinduy}
\affiliation{ESS Bilbao, Poligono Ugaldeguren III, Pol. A, 7B. 48170 Zamudio, Bizkaia - Pa\'{\i}s Vasco, Spain}

\author{T. G.  Perring}
\email[]{toby.perring@stfc.ac.uk}
\affiliation{ISIS Facility,
        STFC Rutherford Appleton Laboratory,
        Harwell Campus, Didcot OX11 0QX, United Kingdom}
\affiliation{London Centre for Nanotechnology, 17-19 Gordon Street, London WC1H 0AH, United Kingdom}

\date{\today}

\begin{abstract}
 The {\sc Horace} suite of programs has been developed to work with large multiple-measurement data sets collected from time-of-flight neutron spectrometers equipped with arrays of position-sensitive detectors. The software allows exploratory studies of the four dimensions of reciprocal space and excitation energy to be undertaken, enabling multi-dimensional subsets to be visualized, algebraically manipulated, and models for the scattering to simulated or fitted to the data. The software is designed to be an extensible framework, thus allowing user-customized operations to be performed on the data. Examples of the use of its features are given for measurements exploring the spin waves of the simple antiferromagnet RbMnF$_{3}$ and ferromagnetic iron, and the phonons in URu$_{2}$Si$_{2}$.
\end{abstract}

\maketitle

\section{Introduction}\label{s:intro}

Neutron spectrometers at central facilities around the world are routinely used to measure the wave-vector, $\mathbf{Q}$, and energy, $\hbar \omega$, dependency of the spectrum of lattice dynamics and magnetic excitations, $S(\mathbf{Q},\omega)$. These data can provide detailed information about the strength, range and symmetry of the interatomic and magnetic interactions, and consequently are highly sensitive tests of theoretical models. The triple-axis spectrometer (TAS) at research reactors has traditionally been the instrument of choice because of its controllability and flexibility \cite{Tranquada-book}, whereby the $S(\mathbf{Q},\omega)$-dependency is explored point-by-point. Over the past 15--20 years time-of-flight spectrometers with position-sensitive detectors (PSDs) have established themselves as extraordinarily effective instruments for measuring excitations in single crystals where the interactions are strong in one or two dimensions, for example in the cuprate and iron-based superconductors \cite{Xu-NPhys,Vignolle-NPhys,Hayden-Nature,Tranquada-Nature,Ewings-Sr122,Zhao-Ca122,Lumsden_PRL09,Lipscombe_ARCS_single,Chan_HgBCO,Song-BFNA-detwin,Zhang-pnictogen-ht,LSCO-4seasons}
and quasi one- and two-dimensional model magnetic systems \cite{Schmidiger,Lake-NMat,Lake-NPhys,Christensen25092007,Notbohm_MAPS,DallaPiazza,CuGeO3-4seasons}. However, until recently there have been relatively fewer measurements in systems where there are significant interactions in all three spatial dimensions. By combining many separate runs, each with a different crystal orientation, into a single data set, complete measurements of the four-dimensional scattering function $S(\mathbf{Q},\omega)$ can be made. This has become possible through the combination of the latest instruments with large solid angle position sensitive detector arrays \cite{MAPS-tech,Bewley20061029,Abernathy-2012,4seasons,Ollivier-IN5} and, crucially, optimized software to visualize and manipulate the massive data sets that are created.

Here the software application {\sc Horace} is described, which is in routine use at several neutron facilities and by their users for the visualization and analysis of such data sets. This paper describes the background to the experimental method, and the principles of {\sc Horace} are outlined. The features of the software are described in detail, together with a summary of how it is practically used, with examples that illustrate its operation and features. Details of computer hardware requirements, download and installation are also summarized.


\section{Theory}\label{s:theory}

\begin{figure}[h]
\includegraphics*[scale=0.6,angle=0]{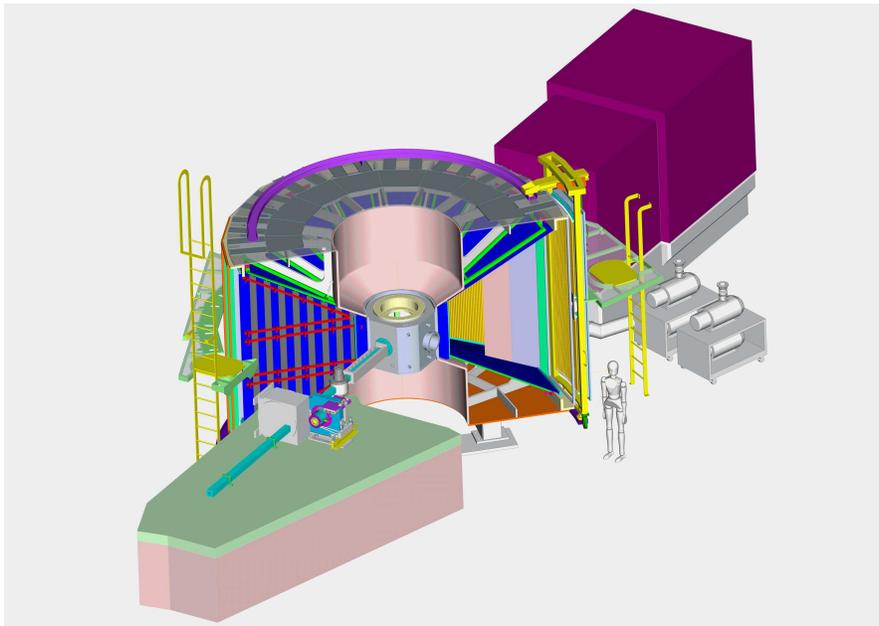}
\centering \caption{Schematic of the {\sc Merlin} chopper spectrometer, at the ISIS spallation neutron source. A white beam of neutrons from the source moderator are incident from the bottom left in this schematic. The principles of operation are described in the text. Such instruments are ideally suited for the technique of combining multiple datasets, with their high flux and large solid angle detector coverage enabling rapid surveys of the 4D scattering function $S(\mathbf{Q},\omega)$ to be undertaken.} \label{fig:Merlin_1}
\end{figure}

Fig. \ref{fig:Merlin_1} shows a schematic diagram of the {\sc Merlin} spectrometer \cite{Bewley20061029} at the ISIS spallation neutron source at the STFC Rutherford Appleton laboratory in the UK, an example of the latest generation of direct geometry spectrometers.  In this example, a pulse of protons hits the spallation target every 20\,ms to produce a pulse of neutrons. These are rapidly slowed down in a moderator to produce a pulse of neutrons with characteristic width measured in microseconds, but with a spread of useable energies in the instrument of $\sim10$\,meV --  $\sim3$\,eV. A monochromatic pulse of neutrons with the desired energy $E_{i}$ is selected by correctly choosing the phase of a rotating collimator (Fermi chopper), or system of disk choppers, just before the sample. The sample scatters neutrons, and on {\sc Merlin} these are collected by a three steradian position sensitive detector array. The time of arrival with respect to the proton pulse of each scattered neutron is recorded together with its location on the detector array.
Because the moderator-to-sample distance $x_{1}$ is known, as is the sample-to-detector distance $x_{2}$ for each detector element, the magnitude of the scattered wave vector for each recorded neutron is determined by the time-of-arrival at the detector, $t_{\rm{det}}$:

\begin{equation}\label{eq:kf_tof_def}
k_{f} = \frac{m_{\rm{N}}}{\hbar} \left(\frac{x_{2}}{(t_{\rm{det}} - t_{\rm{samp}})}\right)
\end{equation}

\noindent where $t_{\rm{samp}} = (m_{\rm{N}} \,/ \,\,\hbar)\cdot (x_{1} \,/\,\, k_{i})$, is the time the monochromatic pulse hits the sample, $k_{i}$ is the magnitude of the incident wave vector given by $E_{i}=\hbar^{2} k_{i}^{2} \,/ \,2 m_{\rm{N}}$, and $m_{\rm{N}}$ is the mass of the neutron.
The momentum and energy transferred to the sample are then computed as

\begin{eqnarray}\label{eq:kinematics}
\mathbf{Q} &=& \mathbf{k}_{i} - \mathbf{k}_{f}\nonumber \\
\hbar \omega &=& \frac{\hbar^{2}}{2m_{\rm{N}}}(k_{i}^{2} - k_{f}^{2})
\end{eqnarray}

For a chosen $k_{i}$ and sample orientation, there are three independent degrees of freedom, corresponding to the spherical polar angles $\theta$ and $\phi$ that define the direction of $\mathbf{k}_{f}$, and the time-of-arrival $t_{\rm{det}}$ which in turn has a one-to-one correspondence with energy transfer, $\hbar \omega$, or equivalently with $k_{f}\equiv|\mathbf{k}_{f}|$. In consequence, the momentum and frequency dependent scattering function $S(\mathbf{Q},\omega)$ is measured on a 3D manifold in the four dimensions of $\mathbf{Q}$ and $\omega$, with the volume defined by the ranges of $\theta$ and $\phi$ set by the size of the detector array, and $-\infty \leq \hbar \omega \leq E_{i}$. Equivalently, in any particular choice of coordinate frame for $\mathbf{Q}$, then of the four coordinates $\{Q_{\alpha}\}\equiv (\mathbf{Q},\omega)$ with $Q_{\alpha}=Q_{\alpha}(\theta,\phi,t_{\rm{det}})$, $\alpha =1-4$, only three components are independent, with the fourth an implicit function of the other three. We note that a similar line of reasoning can be used for indirect geometry spectrometers, for which the final energy $E_{f}$ is fixed and the time-of-flight is used to determine $k_{i}$.

The physically relevant coordinate frames in which to express the components of $\mathbf{Q}$ are ones that are fixed with respect to the crystal lattice. For example, one may choose the components along the reciprocal lattice vectors $a$*, $b$* and $c$*. A good choice of coordinate frame and of which component of $\{Q_{\alpha}\}$ is the implicit coordinate will depend on the material being studied. For example, in some magnetic materials such as the parent high temperature superconductor compound La$_{2}$CuO$_{4}$ \cite{Headings_PRL10}, the magnetic ions are arranged in layers, with the magnetic exchange parameters between the layers orders of magnitude weaker than those within the layers. In this case, the best choice of coordinate frame is one with components $Q_{1}$ and $Q_{2}$ within the layers and $Q_{3}$ perpendicular to the layers. Because the interactions between the layers are negligible, $S(\mathbf{Q},\omega)$ has negligible dependence on $Q_{3}$. In this instance, $Q_{3}$ is taken to be the implicit variable, and the intensity as a function of $(Q_{1},Q_{2},\omega)$ gives the relevant information of $S(\mathbf{Q},\omega)$. Typical plots of $S(\mathbf{Q},\omega)$ at a constant energy are thus projected along the physically uninteresting $Q_{3}$-axis. In quasi-1D magnets, where the interactions are strong only along one direction - label it $Q_{1}$ for definiteness - then $Q_{3}$ can be ignored as the implicit variable and the intensity integrated along $Q_{2}$ to improve the statistical quality of the data, and intensity as a function of $\{Q_{1},\omega\}$ gives the full information of $S(\mathbf{Q},\omega)$. These techniques have been used for many years to study quasi-1D and quasi-2D materials \cite{Xu-NPhys,Vignolle-NPhys,Hayden-Nature,Tranquada-Nature,Ewings-Sr122,Zhao-Ca122,Lumsden_PRL09,Lipscombe_ARCS_single,Chan_HgBCO,Song-BFNA-detwin,Zhang-pnictogen-ht,LSCO-4seasons,Schmidiger,Lake-NMat,Lake-NPhys,Christensen25092007,Notbohm_MAPS,DallaPiazza,CuGeO3-4seasons}, and there are established software applications to visualize the data \cite{Mslice_ref,DAVE_ref}. Though it is possible to use the same techniques and software tools to analyze data from 3D materials \cite{Perring_manganite,Hayden_CrV}, such analysis is far from routine.

In the case of {\sc Merlin} the number of detector elements is $\approx 70,000$, which is is typical of the number for similar instruments at other sources, and the energy transfer axis is typically divided into $\approx 200$ energy bins, so that the 3D manifold is divided into $O(10^{7})$ voxels. Typically there will be six 8-byte numbers associated with each voxel -- $\{Q_{\alpha} \}$, $\alpha=1-4$, intensity and error on the intensity - which accounts for the bulk of any representation of the data set, which in the case of MERLIN amounts to $\approx 0.5$\,GBytes. This is sufficiently small to fit easily into the memory of a commodity PC.

To map $S(\mathbf{Q},\omega)$ fully in materials with interactions in all three spatial directions requires an extra degree of freedom. A sequence of data sets is collected, in which an additional parameter is successively incremented between each data collection, or `run', and the entire collection of data sets is treated as one. Two approaches are possible for use with {\sc Horace}:

\begin{enumerate}
 \item Rotate the sample about an axis (usually vertical) by some angle, $\Psi$, typically $0.5^{\circ} - 2^{\circ}$ between each run, while keeping the incident neutron energy $E_{i}$ fixed (in the case of an indirect geometry spectrometer the final neutron energy $E_{f}$ is fixed). The range of the angular scan is usually determined by the reciprocal space coverage of the instrument's detectors and the symmetry of the crystal lattice. This method of operation is the one most frequently used.

\item Keep the sample orientation fixed, but increment the incident energy $E_{i}$ by a small amount between runs. This mode is very rarely used, since the scan range in $E_{i}$ must be kept small in order to avoid too much variation in the instrumental resolution between runs, and this results in rather limited reciprocal space coverage. For the following we shall ignore this option in our description of the method, and focus on sample rotation.
\end{enumerate}

Although the program does allow runs with different orientation \emph{and} incident energy to be combined, this is generally not advisable since this can result in a multi-valued resolution function for a given $(\mathbf{Q},\omega)$ voxel with different $E_{i}$ and $\psi$, making interpretation and analysis or the data much more difficult.

With these choices $Q_{\alpha}=Q_{\alpha}(\theta,\phi,t_{\rm{det}},\Psi)$ or $Q_{\alpha}(\theta,\phi,t_{\rm{det}},E_{i})$,  $\alpha=1-4$, the scattering function $S(\mathbf{Q},\omega)$ is measured in a 4D manifold. The choice of increment in the additional parameter is based on consideration of the resolution of the instrument. The angular divergence of both the incident and scattered neutron beams is $0.5^{\circ} - 1^{\circ}$ for current chopper spectrometers, and the energy resolution is typically $\Delta \hbar \omega \, /\,E_{i} = 1 - 6\%$. With continuous streaming of data to disk as a function of absolute time rather than time relative to the most recent proton pulse (that is, event mode collection), then $\Psi$ can in principle be varied continuously, avoiding discretization of data along the corresponding coordinate axis \cite{Utsusemi-ref,Weber_ARCS_continuous}. In practice, however, runs at different $\Psi$ usually are discretized in order to simplify matters in case of equipment failure part-way through a set of measurements, so that `bad' runs may be easily eliminated.


\section{Program Description}\label{s:prog}

\subsection{Main Purpose}\label{ss:purpose}

The main purpose of {\sc Horace} is to allow easy visualization, manipulation and analysis of inelastic neutron scattering data, gathered from multiple crystal orientations or incident energies as described above, in the four dimensions of vector momentum and energy. {\sc Horace} provides a comprehensive set of elemental functions as an extensible framework for analyzing the data, in which more complex scripts or functions can be rapidly written by users. A guiding principle in the design of {\sc Horace} was that it should be possible to run on a high specification laptop or commodity desktop computer, and would pre-process the data to minimize the time to create and visualize subsets cut from the data, despite that hardware constraint.

\subsection{Architecture and key features}\label{ss:architecture}

{\sc Horace} has been written with an object oriented architecture in the commercial high-level technical programming language Matlab \cite{Mex-ref}. The primary object is the {\it sqw} object: this contains the signal and variance for each individual detector-energy voxel, meta information describing the detector locations, crystal lattice and crystal orientation for each contributing measurement, and the mapping of the voxels into a Cartesian grid that defines the current plot axes and bin sizes in one, two, three or four dimensions. In addition there is the {\it dnd} object: an abbreviated version of the sqw object that does not retain the information of the individual voxels or the detector locations. Generally a dnd object will occupy several orders of magnitude less computer memory than the equivalent sqw object.

The operations that are supported on sqw and dnd objects include:

\begin{itemize}
\item Construction of 4D sqw data from multiple measurements of inelastic neutron scattering data.
\item Creating new sqw objects by taking 1D, 2D, 3D and 4D sub-manifolds from the original sqw object, or any other sqw object created by cutting from a previously created sqw object. We define these as cuts; the axes of the cuts can be chosen to be in arbitrary directions in momentum space, or energy.
\item Plotting 1D, 2D and 3D objects.
\item Unary operations e.g. correction for the detailed balance factor, as well as the standard operations such as sign inversion and trigonometrical functions.
\item Binary operations +, -,$\ast$, $/$, $^{\wedge}$.
\item Replication of a lower dimensional cut along the additional axes of a higher dimensional cut . This is useful e.g. for creating background estimates to be subtracted from the higher dimensional cut.
\item Symmetrization of sqw objects.
\item Simulation and fitting of models of the scattering function $S(\mathbf{Q},\omega)$.
\end{itemize}

In addition, there is a tool for planning the range of crystal orientations at which to make measurements in order to map a desired volume of momentum and energy, and utilities for plotting data or models of dispersion relations and the scattering function as a function of energy and momentum along a sequence of high-symmetry directions in reciprocal space. Examples of the use of these operations on illustrative data sets will be given in Section \ref{s:use}.

In order to satisfy the requirement that {\sc Horace} can operate on a commodity personal computer and to minimize the time to make a cut, the data are coarse-grain sorted on to a regular 4D grid in the $(\mathbf{Q},\omega)$-space (by default fifty steps along each dimension), and then saved into a single (generally large) file with the extension {\it .sqw}. The advantage of this sorting when the data are saved to file is that to access a particular volume of reciprocal space and energy window, which is typically the case when visualizing and analyzing a 1D or 2D cut, the entire large sqw file need not be read from disk to search for contributing detector voxels. Instead, just the data in those bins in the grid which intersect the reciprocal space volume and energy window of the cut need to be read. This provides a significant saving of time when extracting such subsets, and is one of the key features of {\sc Horace} compared to its predecessors.


\subsection{Interface}\label{ss:interface}

\begin{figure}[!h]
\includegraphics*[scale=0.7,angle=0]{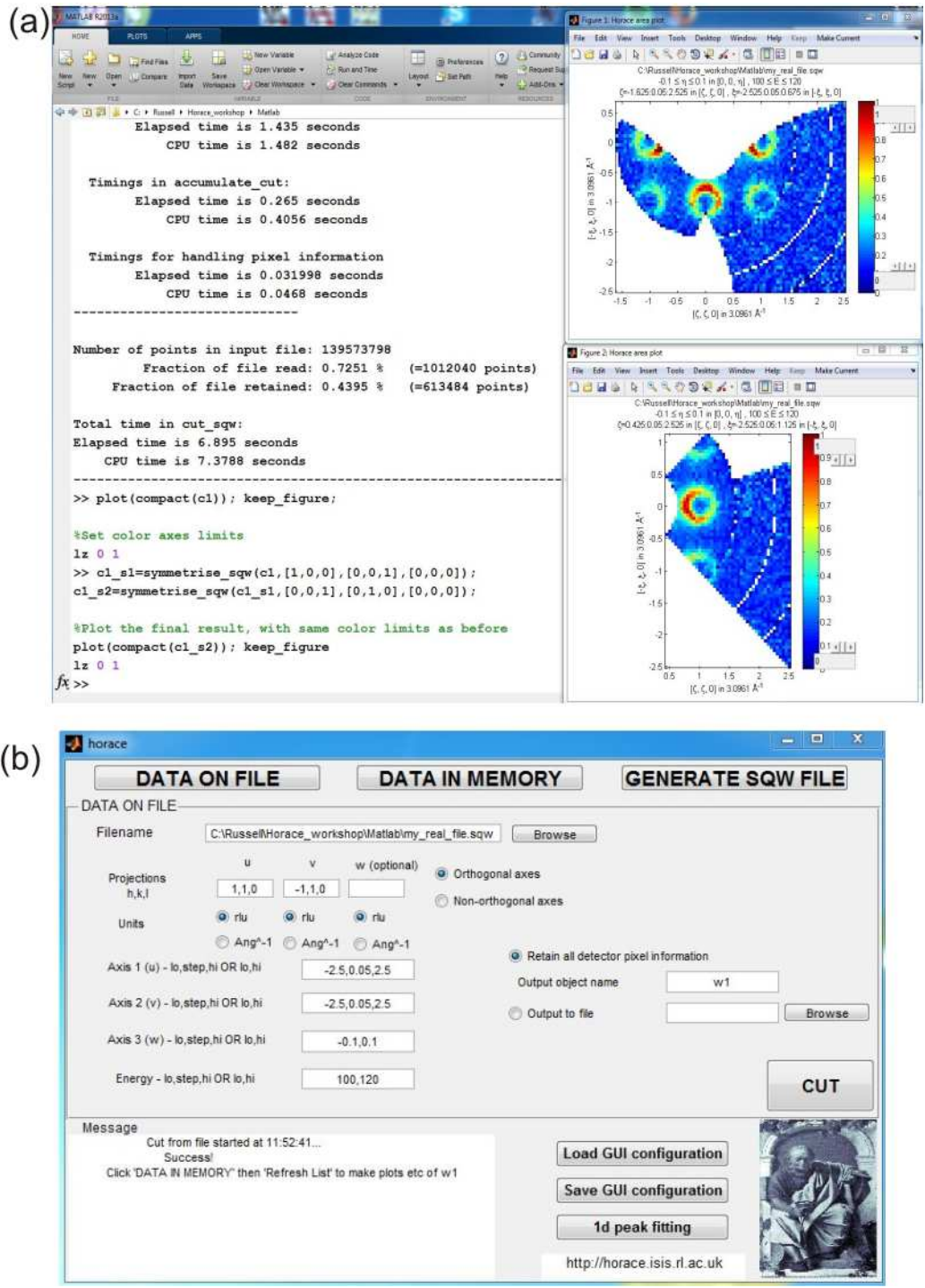}
\centering \caption{(a) screen shot of the Matlab command window, showing a series of commands to extract a 2D cut from a file, plot it, symmetrize in two planes, and finally plot the result. The outputs from the various operations are also shown in the command window, as are the two plots of the unsymmetrized and symmetrized data. (b) screen shot of the {\sc Horace} GUI, with fields filled in to make the same cut as shown in panel (a).} \label{fig:Commandwindow}
\end{figure}

Every operation that {\sc Horace} can perform has been written as a Matlab function. There are two possible ways of interacting with the program. The first, and most common, is through the Matlab command line, or for more complex sequences of commands through the user's own Matlab scripts or functions. An example of {\sc Horace} in use with the Matlab command line is shown in Fig. \ref{fig:Commandwindow}(a), together with two plots generated during the series of commands shown. In this particular case a cut is taken from a file and plotted. The cut is then symmetrized in two planes, and the result of this operation is also plotted. The second method for interacting with {\sc Horace} is through a graphical user interface (GUI), shown in Fig. \ref{fig:Commandwindow}(b), which allows access to a subset of the functionality of the main program chosen to allow execution of the most common and / or simplest tasks. Note that use of the GUI still requires Matlab to be installed on one's computer.


\subsection{Operation}\label{ss:operation}

Practical operation of {\sc Horace} is a two-stage procedure. The first, pre-processing, stage is the creation of the sqw files, which is done for on-the-fly analysis during experiment as data accumulate, and then usually once at the end of an experiment to create reference sqw files for later analysis. These are created from multiple individual run files, which contain the measured scattering intensity $S(\mathbf{Q},\omega)$ for each detector as a function of energy transfer. Such files are created from the raw time-of-flight data, with correction for detector efficiency, absorption by the sample, etc. using, for example, the Mantid software \cite{Arnold2014_Mantid}. Each individual run file is sorted on to a common 4D grid, and then the sorted files are combined a piece at a time into the final sqw file. This file-backed combination limits the memory that is required, which is crucial when one considers the size of the final sqw file, which can typically range in size from 10 to 500\,GB.

Once the sqw file has been created, cuts of any dimensionality can be taken and any of the manipulations detailed in Sec. \ref{ss:architecture} applied to them. We include in this the ability to plot 1D, 2D, and 3D cuts, since this is the way in which the user typically interacts with the data. A typical workflow might be to take a series of cuts and plot them to investigate some region of interest, apply some corrections (e.g. magnetic form factor, background subtraction, or Bose-Einstein population factor), then simulate and fit a model to these data in order to extract some physically meaningful parameters.


\section{Illustrative Examples - RbMnF$_{3}$, Fe and URu$_{2}$Si$_{2}$}\label{s:examples}

We will give a basic illustration of some of the functionality of {\sc Horace} with reference to data taken on the following three examples, RbMnF$_{3}$, iron and URu$_{2}$Si$_{2}$. RbMnF$_{3}$ has a cubic crystal structure and is very close to being an ideal 3D Heisenberg antiferromagnet. It has a large spin ($S=5/2$) on its Mn$^{2+}$ sites, making it a strong magnetic scatterer of neutrons. It has a nearest-neighbor isotropic exchange constant of $J=0.29$\,meV, and a next-nearest-neighbor exchange constant that is an order of magnitude smaller \cite{Windsor-RbMnF3}. Iron is the canonical example of an itinerant-electron metallic magnet. Below about $100$\,meV sharp spin waves have been shown to exist \cite{Collins-iron}, but time-of-flight inelastic neutron scattering experiments have also shown that spin fluctuations persist up to much higher energies of at least $550$\,meV \cite{Perring-iron}. The data shown here will be the subject of future scientific reports, and are used here for illustrative purposes only. URu$_{2}$Si$_{2}$ has been actively studied for many years due to the mysterious `hidden order' that it exhibits \cite{Mydosh_rev}, which is responsible for a large change in entropy but cannot be explained by a conventional order parameter such as dipolar magnetic order. Recent interest has focused on whether the lattice is coupled to the hidden order parameter, and several studies of the phonons have been published \cite{Buhot_prb,Butch_prb}. The data we show here are in agreement with the already-published work, but cover a much larger volume of reciprocal space by virtue of the fact that we used the method of data collection outlined in this paper.

\section{Use of the program}\label{s:use}

We now provide a more comprehensive discussion of how the program is typically used.

\subsection{Planning an experiment}\label{ss:planning}

\begin{figure}[h]
\includegraphics*[scale=0.7,angle=0]{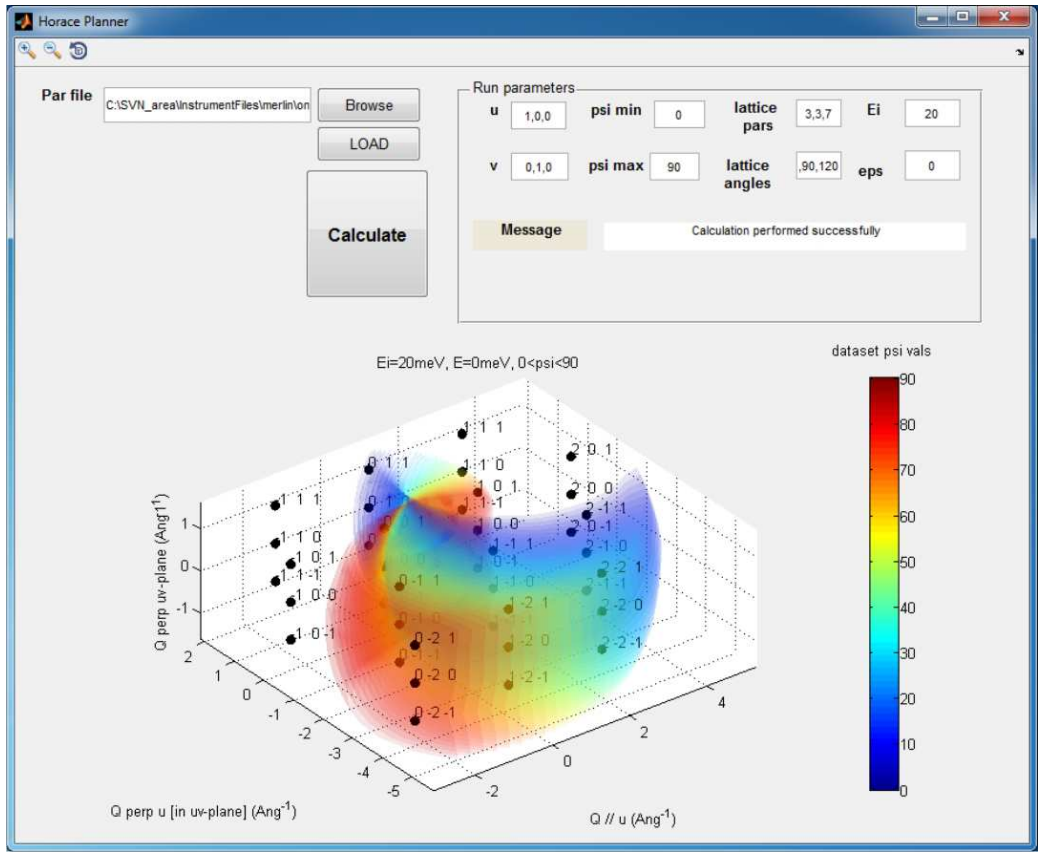}
\centering \caption{Screen shot of the {\sc Horace} scan planner, illustrating how the reciprocal space coverage for a given instrument, incident energy and scan range combination may be calculated in advance of performing the measurements.} \label{fig:Planner}
\end{figure}

The first step when performing an experiment is often to determine an appropriate choice of instrument parameters, such as incident neutron energy and range of sample orientations to be scanned. Different combinations give access to different regions of reciprocal space, and with the limited time available to run an experiment it is crucial to decide quickly the right instrumental configuration. To help with this a standalone GUI is provided with {\sc Horace}, the `scan planner'. Given a set of basic inputs concerning the lattice parameters and angles, sample orientation, instrument detector positions and incident neutron energy, the program plots the volume of $\mathbf{Q}$-space covered by those detectors for a given energy transfer. To aid the planning process the volume is semi-opaque and colored according to the sample orientation angle, and the positions of integer $(H,K,L)$ are plotted as black spheres. An example screenshot is shown in Fig.\,\ref{fig:Planner}.

\subsection{Combining multiple data files}\label{ss:gen_sqw}

The selection of individual run files to combine into an sqw file may be performed either in a Matlab script or through the GUI. Metadata about the experimental setup for each run must be manually provided by the user, specifically the incident energy, sample orientation, lattice parameters and lattice angles. For the RbMnF$_{3}$ data set we combined 85 individual measurements at different sample orientations, each of which was 122\,MB, taken on the MAPS time-of-flight neutron spectrometer at ISIS \cite{MAPS-tech}. The scattering plane was $(1,1,0)/(0,0,1)$ and $\Psi$ was scanned from $6^{\circ}$ to $90^{\circ}$ in $1^{\circ}$ steps. The resultant sqw file was 15\,GB, and additional working space of about the same size on disk as the sqw file was required during its creation. For the iron data set, also obtained on the MAPS spectrometer, 186 runs were combined to make a sqw file of 36\,GB. The scan was performed with the $(1,0,0)/(0,1,0)$ scattering plane, and $\Psi$ scanned from $-92.5^{\circ}$ to $0^{\circ}$ in $0.5^{\circ}$ steps. The URu$_{2}$Si$_{2}$ dataset was obtained on {\sc Merlin}, and comprised 276 runs which combined gave an sqw file of size 136\,GB. Generally speaking a larger number of runs and / or instruments with a larger number of detector elements give rise to larger sqw files, which take commensurately longer to generate.

Speed-up of the creation of the sqw file, and other computationally intensive operations such as taking cuts (see below), is achieved by using C++ routines in place of Matlab ones. These are invoked using Matlab's in-built mex file system \cite{Mex-ref}, whereby Matlab routines may call subroutines written in another language. The C++ routines utilize both multi-threaded processing as well as the intrinsic speed gains that are typically obtained when comparing an interpreted language (Matlab) with a compiled language (C++).

The time taken to perform the combination of files is highly dependent on the computer on which it is performed. For benchmarking we used a sqw file of size 142\,GB, which comprised data from 231 runs. On a Windows 7 workstation with available disk space of several TB (i.e. much more than the sqw file size), with 48\,GB RAM and running 12-core Intel Xeon X5650 processors (2.67\,GHz), the total time to generate the sqw file was 150 minutes when using Matlab 2015b (later versions of Matlab include internal multi-threading procedures, which offer similar speed to Mex acceleration in this case). On a machine running CentOS7 with the same hardware, and with Mex file acceleration enabled and running on 8 threads, the total optimized time was 52 minutes. However, recent versions of Horace contain extensions to the code that allow better utilization of high performance computing capabilities that are increasingly available. By way of example, the {\it ISIScompute} service available to ISIS facility users, which comprises a machine running RHEL 7 with 96 Intel Xeon E5-4657L processors (2.5\,GHz), 512\,GB of RAM and a 100\,TB CEPH parallel file system \cite{CEPH-ref}, is able to produce the same 142\,GB file in around 8 minutes.

We have noted already that {\sc Horace} has been designed to be operable on a typical commodity PC, which may well have a lower specification than that described above for our benchmarks. Provided sufficient disk space is available to store the sqw file and the temporary working space needed during its creation, lack of RAM and CPU speed need not prevent the operation of {\sc Horace}. Options are provided whereby the size of chunks read from disk to memory for processing can be changed, so for a PC with less RAM these numbers can be reduced appropriately. The time taken to generate files and take cuts from them will increase, and the number of sqw objects that can be held in memory is smaller, but otherwise the full functionality of {\sc Horace} is available.

During a typical experiment the user will often wish to examine data from a partially complete scan of sample orientations, in order to make decisions about what future measurements to make. Rather than regenerating the entire sqw file when more runs have been completed, it is possible to provide a list of \emph{planned} runs and sample orientations, so that future data may be binned on to the same coarse-grained grid and inserted into the existing sqw file, thus saving time, especially with larger files.


\subsection{Extracting and visualizing data}\label{ss:visualisation}

Once the sqw file has been created, {\sc Horace} is also used to visualize and analyze the data. Typically users wish to sample 3D volumes, 2D slices and 1D cuts along specified trajectories in $(\mathbf{Q},\omega)$-space. As mentioned in Sec. \ref{ss:architecture}, we refer to such subsets in general terms as `cuts'. {\sc Horace} provides complete flexibility to cut along any $\mathbf{Q}$-direction, or along the energy direction, irrespective of the orientation of the sample with respect to the instrument. When making a cut the user specifies a grid onto which data are binned, with the bin sizes chosen typically to contain data from many detector voxels. It is at this stage that the coarse grain sorting of the data during the generation of the sqw file, described earlier, provides a significant speed advantage since only a small fraction of the total (very large) file needs to be read to obtain all the information required for any given cut. Once these cuts from the data are read from disk they are stored in memory and are accessible as objects in the Matlab `workspace', so that provided sufficient computer memory is available multiple cuts may be retained for future visualization and/or analysis. Every cut has the same structure as the data in the sqw file, so further cuts may be taken from objects in memory without loss of information.

\begin{figure}
\includegraphics*[scale=0.6,angle=0]{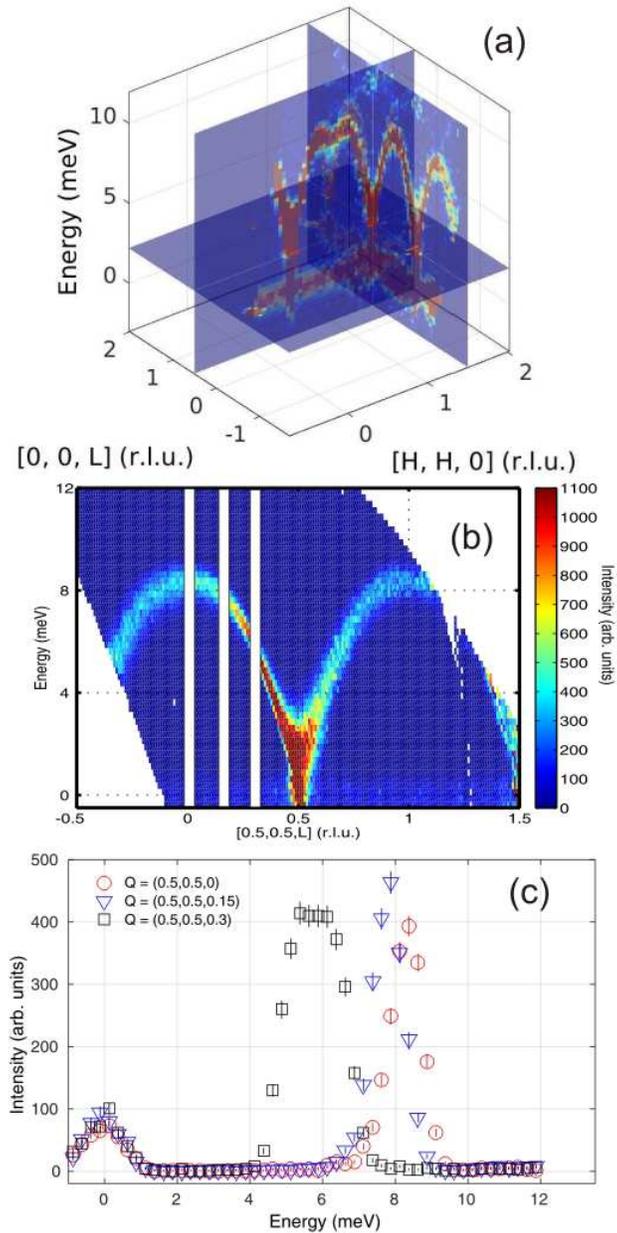}
\centering\caption{Plots of data from the experiment on RbMnF$_{3}$. Panel (a) shows a screen-shot of the sliceomatic feature of {\sc Horace}, which allows visualization of 3D cuts. Panel (b) shows a 2D slice in the $(0.5,0.5,L)$ - Energy plane. The white lines show regions where 1D cuts were taken. Panel (c) shows three 1D cuts for $L=0$ (red circles), $L=0.15$ (blue triangles), and $L=0.3$ (black squares).}
\label{fig:Multipane_RbMnF3}
\end{figure}

{\sc Horace} provides tools to visualize 1D, 2D and 3D cuts as, respectively, marker plots with errorbars, colormaps, and multiple colormaps plotted on a 3D set of axes, examples of which are shown in Fig.\,\ref{fig:Multipane_RbMnF3}. These plots are highly customizable, and because the plots are ultimately generated using Matlab's native graphics they can also be modified using the in-built routines. It is thus fairly common that {\sc Horace} is used directly to produce figures that are used in publications, in addition to being used for the analysis \cite{Mena_prl,KimJain_prl,Parshall_prb,Inosov_ARCS_horace,Plumb_SEQ_horace,SEQ_Fuhrman_horace,SEQ_Taylor_horace,Baker_IN5_Horace,Toth-PRL,Jeong-PRL,Tomiyasu-PRL}

In order to survey a large section of reciprocal space (i.e. a 3D cut) the sliceomatic tool \cite{Sliceomatic-ref} is used. A screen-shot of sliceomatic in use is shown in Fig.\,\ref{fig:Multipane_RbMnF3}a, with the intensity of the scattering given by a color map. Dispersive excitations with the same periodicity as the Brillouin zone are clearly visible. The user may move the visible slice planes on this interface, in order to explore a large section of the data very quickly. Fig.\,\ref{fig:Multipane_RbMnF3}b shows a 2D slice,
centered on the $(1/2,1/2,1/2)$ position with axes of $(1/2,1/2,L)$ and neutron energy transfer. The slice clearly shows scattering from a band of dispersive magnetic excitations in the range $0\leq E \leq 9$\,meV. The white lines at $L=0$, $L=0.15$, and $L=0.3$ show where 1D cuts were made -- these cuts are shown in Fig.\,\ref{fig:Multipane_RbMnF3}c. The cuts were taken by averaging the signal along $L$, $\pm 0.05$\,r.l.u. either side of the stated value.

\begin{figure}[h]
\includegraphics*[scale=0.7,angle=0]{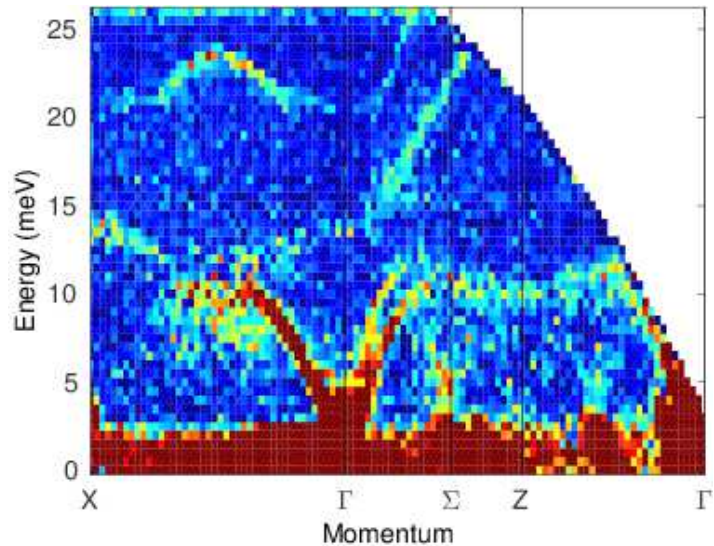}
\centering \caption{`Spaghetti plot' of the phonon dispersion in URu$_{2}$Si$_{2}$ around $\mathbf{Q} = (2,-2,0)$. The high symmetry points are indexed with respect to the body-centered tetragonal Brillouin zone, which in the simple tetragonal notation usually used for this material are $\Gamma = (0,0,0)$, $\Sigma = (0.6,0,0)$, $Z=(0,0,1)$, $X=(1,1,0)$.} \label{fig:spaghetti}
\end{figure}

It can often be useful to view the dispersion along several high symmetry directions on a single plot, for example when investigating phonons and comparing to DFT calculations. The Horace tool \texttt{spaghetti\_plot} can be used for this purpose. An example of its use is shown in Fig.\,\ref{fig:spaghetti} in which we show the phonon dispersion around $\mathbf{Q}=(2,-2,0)$ in URu$_{2}$Si$_{2}$. One can see, for example, the splitting of two different acoustic modes along the $\Gamma$ -- $\Sigma$ trajectory as well as multifarious modes at higher energies which disperse differently along different symmetry directions.


\subsection{Manipulating data}\label{ss:manipulation}

There are several different ways one can manipulate sqw and dnd objects. Unary operations that apply to the intensity, e.g. Bose-Einstein population factor or magnetic form factor correction of the intensity, and binary operations, e.g. subtraction of the intensity of one object from another (such as required for background subtraction), may be performed.

Data of dnd form may be smoothed by convolution with an appropriate dimensional Gaussian or hat function of a specified width. Such smoothed objects allow a simple way to visualize data part-way through an experiment, before sufficient statistical quality has been obtained through longer measurement times. By definition sqw data may not be smoothed, since the relationship between the intensity that is plotted and the underlying detector voxel information must be maintained for such objects.

It is possible to repeatedly tile a lower dimensional dnd data set into a higher dimensional one, e.g. replicate a 1D cut along the energy axis along some $\mathbf{Q}$-axis to create a 2D cut with axes of $\mathbf{Q}$ and energy. This is useful when performing background subtractions, since in some cases the intrinsic background may depend on energy but weakly or not at all on $\mathbf{Q}$. Thus a 1D cut along energy may be taken in some region where there is only background, and then this can be subtracted from another region of the data to leave just the contribution to the signal from the intrinsic $S(\mathbf{Q},\omega)$. An example of such a procedure is shown below in Fig.\,\ref{fig:Fe} for the iron dataset. Here, a region (highlighted by the dashed rectangle) is selected that is representative of the non-magnetic background and a 1D cut is performed (panel b). This cut is then replicated over the full Q-range of the original 2D slice (panel a) and then subtracted (panel c).

\begin{figure}
\includegraphics*[scale=0.7,angle=0]{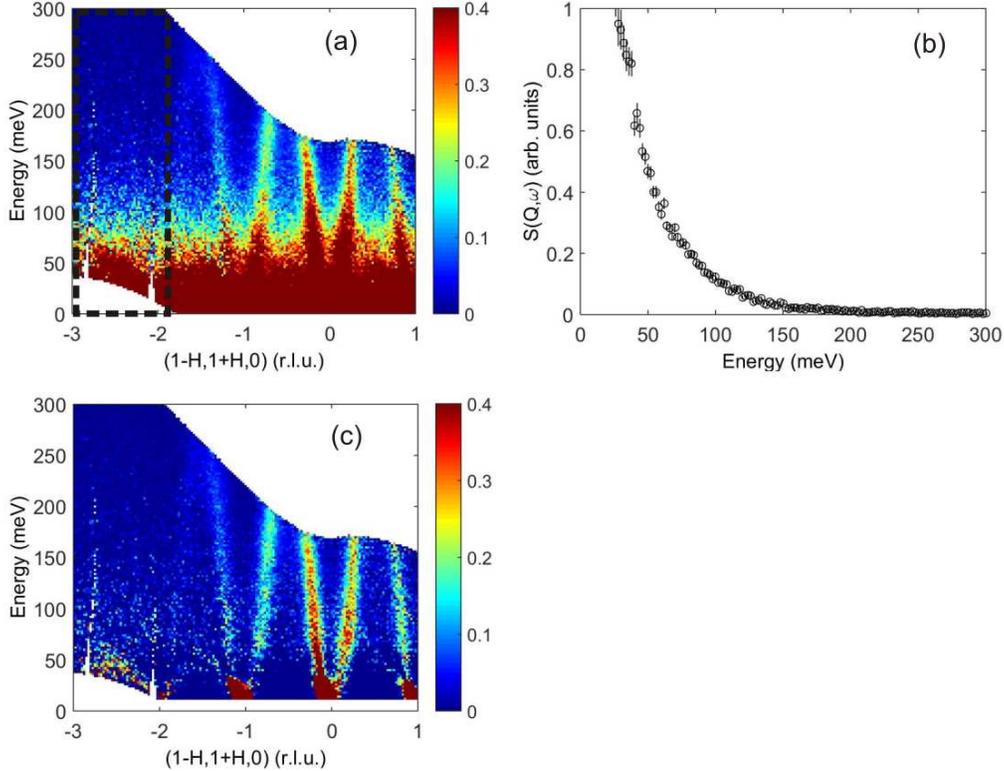}
\centering \caption{(a) Representative $(\mathbf{Q},E)$-slice showing dispersive magnetic excitations in iron, together with incoherent elastic and phonon scattering. The dashed rectangle shows an area expected to have negligible magnetic scattering and hence representative of the non-magnetic background. (b) Cut through the non-magnetic region in the dashed rectangle. (c) Result of replicating the 1D background cut and then subtracting this from the original data.} \label{fig:Fe}
\end{figure}


\subsection{Simulating and fitting}\label{ss:simandfit}

{\sc Horace} provides the ability to fit and simulate (which is simply a single evaluation of a fit function) the data for precisely the same values of $(\mathbf{Q},\omega)$ that were measured. Models for fitting can take two forms, either generic functions of the plot coordinates (e.g. Gaussian peaks and the like) or more physically meaningful models that calculate $S(\mathbf{Q},\omega)$ directly. The former is useful for fitting, for example, peak functions on 1D cuts to give a quick parameterization of a dispersion relation. The latter are much more powerful, and can be used to determine physical parameters directly from the data.

It is particularly for the fitting of $S(\mathbf{Q},\omega)$ models that the full detector voxel information retained in sqw objects is most useful. The model function is evaluated for all of the voxels, and then combined to give the intensity in a particular $(\mathbf{Q},\omega)$-space bin, rather than just at the bin center as would be the case for a dnd object. For models where $S(\mathbf{Q},\omega)$ varies appreciably across the width of one bin, this can result in systematic problems if evaluating only using dnd objects, whereas sqw objects generally provide a more accurate fit. On the other hand, because there are usually many detector voxels contributing to the signal in a given bin, evaluation of a model using an sqw object can take much longer due to the greater amount of computer processing required.

The fitting routines provided with {\sc Horace} are designed to work to a high level of abstraction. A key feature of the fitting capability of {\sc Horace} is that fits can be performed on an arbitrary number of cuts of any dimensionality, using a global model with global parameters. The fit routines allow a distinction between `foreground' (often global) and `background' (usually local to each cut), with the former typically being a model of $S(\mathbf{Q},\omega)$ and the latter being generic function(s) such as a linear sloping background. This distinction is especially powerful when fitting multiple cuts, since a larger part of the overall dataset can be used to constrain a model, achieving higher accuracy, while allowing for the fact that the instrumental background often varies in unusual ways from cut to cut. Such a philosophy for the fitting was developed in recognition of the form that real data take. As is the case for most fitting procedures, one can specify which fit parameters can vary or remain fixed, and can also bind parameters together in a fixed ratio. Fit functions can take as inputs information of any form (e.g. lookup tables, as well as numeric parameters). Instrumental resolution broadening can at present be included in a crude way as part of the fit function, such a applying a Gaussian broadening in energy, but a specific model for instrumental resolution is not included in {\sc Horace} at the moment.


\section{Summary}\label{s:summary}

We have written a suite of programs, {\sc Horace}, to take multiple runs from time-of-flight neutron inelastic scattering experiments, and combine them in one single large data set that can be hundreds of GB in size. The program is designed to be an extensible framework that allows a range of sophisticated manipulations to be performed on the data. The program is also used to visualize subsections of the large data set, with a coarse grained sorting of detector voxels' $(\mathbf{Q},\omega)$ coordinate ensuring fast access to the relevant subsection of the large data file. The program may also be used for simulations and fits to the data with $S(\mathbf{Q},\omega))$ models. This includes the ability to fit multiple datasets with a global foreground model and set of parameters, but independent background models and parameters. This is a method geared towards the physical origin of the measured signal, and provides a convenient framework for performing the kind of analysis which is often done in an {\it ad-hoc} fashion otherwise.


\section{Distribution and documentation}\label{s:distribution}

Other than sufficient disk space to store the spe and sqw files, whose sizes are rather dependent on the instrument used for the measurements, the main hardware requirement is to have at least 8\,GB of RAM. It has been found that less than this severely hampers the user's ability to exploit {\sc Horace} fully with their data. {\sc Horace} has been tested on the following operating systems: 32-bit and 64-bit Microsoft Windows, 64-bit RHEL 6 and 7, and Ubuntu Linux 10.04 and later, and on Mac OS X 10.5.6 and later. {\sc Horace} is written using Matlab, and has been tested on Matlab versions 2009a onwards. {\sc Horace} will continue to be supported in the future for at least the most recent five years' worth of Matlab versions. {\sc Horace} is actively maintained for the above operating systems, but in principle, provided one is able to run a sufficiently recent version of Matlab, it should be possible to run {\sc Horace} no matter what the operating system (e.g. other Linux distributions) without Mex files. The C++ code is also available for the user to perform their own compilation of Mex files if desired.

Zip files containing the compiled {\sc Horace} Matlab code can be downloaded from \url{http://horace.isis.rl.ac.uk}. The full source code is available on request. Users are requested to register an email address when they download the code from the website, so that they can be informed from time to time of new releases and bug-fixes. Installation involves simply unzipping the download into a suitable directory, adding this directory to the Matlab path, and running a short Matlab routine called \texttt{horace\underline{{ }}on} to initialize a more complete and self-consistent set of search paths for Matlab. A full manual, giving complete instructions on installation and use, is also available at this website.

\section{Acknowledgements}\label{s:ack}
We are grateful to Ross Stewart, Tatiana Guidi, Rob Bewley, Helen Walker, Devashibhai Adroja, David Voneshen, and the many users of {\sc Horace} worldwide for comments and feedback about the program. We thank H. A. Mook for the loan of the iron crystal, and W. J. L. Buyers for the loan of the URu$_{2}$Si$_{2}$ crystal. Experiments at the ISIS Pulsed Neutron and Muon Source were supported by a beamtime allocation from the Science and Technology Facilities Council.

\bibliography{refs_horace_paper}

\end{document}